\documentclass[12pt,a4paper]{article}
\pdfoutput=1
\usepackage{amsmath}
\usepackage{diagbox}
\usepackage{jheppub}

\DeclareMathOperator{\Gr}{Gr}

\title{\Large Cluster Adjacency for $m=2$ Yangian Invariants}

\author[1]{Tomasz \L ukowski,}\emailAdd{t.lukowski@herts.ac.uk}
\author[2]{Matteo Parisi,}\emailAdd{parisi@maths.ox.ac.uk}
\author[3,4]{Marcus Spradlin}\emailAdd{marcus\_spradlin@brown.edu}
\author[3]{and Anastasia Volovich}\emailAdd{anastasia\_volovich@brown.edu}

\affiliation[1]{School of Physics, Astronomy and Mathematics, \\
 University of Hertfordshire, \\
 Hatfield, Hertfordshire, AL10 9AB, UK}

\affiliation[2]{Mathematical Institute, University of Oxford, \\
 Andrew Wiles Building, Radcliffe Observatory Quarter, \\
 Woodstock Road, Oxford, OX2 6GG, UK}

\affiliation[3]{Department of Physics, \\
 Brown University, \\
 Providence, RI 02912, USA}

\affiliation[4]{Brown Theoretical Physics Center, \\
 Brown University, \\
 Providence, RI 02912, USA}

\abstract{We classify the rational Yangian invariants of the $m=2$ toy model of $\mathcal{N}=4$ Yang-Mills theory in terms of generalised triangles inside the amplituhedron $\mathcal{A}_{n,k}^{(2)}$. We enumerate and provide an explicit formula for all invariants for any number of particles $n$ and any helicity degree $k$. Each invariant manifestly satisfies cluster adjacency with respect to the $\Gr(2,n)$ cluster algebra.}

\begin{document}

\maketitle


\section{Introduction}

Aspects of the $\Gr(4,n)$ Grassmannian cluster algebras~\cite{FZ1, GSV, scott2006grassmannians} have been found to play several still rather mysterious roles in the mathematical structure of scattering amplitudes in planar $\mathcal{N}=4$ Yang-Mills theory~\cite{Golden:2013xva, Golden:2014xqa}. A simple toy model which serves as a nice playground for studying features of this cluster structure is the $m=2$ version of the theory, where the momentum twistors describing the kinematic scattering data~\cite{Hodges:2009hk} are restricted to lie in a $\mathbb{P}^1$ subspace\footnote{Note that this is quite different from restricting to two space-time dimensions.} of the usual $\mathbb{P}^3$. The associated $\Gr(2,n) \cong A_{n-3}$ cluster algebra\footnote{This algebra has also been found to govern the structure of $\mathcal{N}=4$ Yang-Mills amplitudes in the multi-Regge limit~\cite{DelDuca:2016lad}.} is completely understood~\cite{FZ1}: its clusters are in one-to-one correspondence with the triangulations of an $n$-gon, and the coordinates in each cluster are in one-to-one correspondence with edges in the corresponding triangulation. The positive geometry \cite{Arkani-Hamed:2017tmz} associated to these $m=2$ amplitudes is the amplituhedron $\mathcal{A}_{n,k}^{(2)}$, which is an interesting object on its own, and its many interesting features were studied e.g.~in~\cite{karp2017decompositions,Galashin:2018fri,Ferro:2018vpf,Lukowski:2019kqi}.

In this paper we explore the conjecture~\cite{Drummond:2018dfd,Mago:2019waa} that the poles of every rational Yangian invariant are given by cluster coordinates, a property referred to as \emph{cluster adjacency} following~\cite{Drummond:2017ssj}. The full $\mathcal{N}=4$ Yang-Mills theory has a whole zoo of $n$-particle N${}^k$MHV Yangian invariants (see Chapter~12 of~\cite{ArkaniHamed:2012nw} for a discussion of their classification), and evidence supporting this conjecture is so far restricted to relatively small $n$ and $k$. In contrast, in the $m=2$ toy model, by using the amplituhedron formulation of scattering amplitudes~\cite{Arkani-Hamed:2013jha}, we are able to write down an explicit formula for all Yangian invariants for any $n$ and $k$. Each N${}^k$MHV invariant is labelled by a collection of $k$ non-intersecting triangles inside an $n$-gon, with denominator factors corresponding precisely to edges of the triangles. Consequently, the result manifestly satisfies cluster adjacency with respect to the $\Gr(2,n)$ cluster algebra.

\section{Classification of Yangian Invariants for \texorpdfstring{$m=2$}{m=2}}
\label{sec:2}

\emph{Yangian invariants} are basic building blocks for many amplitude-related quantities of interest (see for example~\cite{Mason:2009qx, ArkaniHamed:2009vw, ArkaniHamed:2009dg, ArkaniHamed:2009sx, Drummond:2010uq, Ashok:2010ie, ArkaniHamed:2012nw, Drummond:2010qh,Ferro:2016zmx}).
The classification of $\mathcal{N}=4$ Yang-Mills invariants (i.e., $m=4$) is discussed in Sec.~12 of~\cite{ArkaniHamed:2012nw}. In this section we discuss the classification of the analogous set of Yangian invariants for $m=2$. We will see that they can all be associated to the so-called generalised triangles, which are the building blocks for triangulations of the amplituhedron space $\mathcal{A}_{n,k}^{(2)}$. Then the Yangian invariants can be extracted from canonical differential forms with logarithmic singularities on all boundaries of generalised triangles.  For $k=1$  there is a unique type of Yangian invariants of the form~\eqref{omegaabc} which trivially corresponds to a triangle in $\mathbb{P}^2$~\cite{ArkaniHamed:2010gg}. For $k=2$ there are two types of Yangian invariants, \eqref{omegaabcabc} and~\eqref{omegaabcd}, which we will see correspond respectively to two non-intersecting triangles or to two triangles glued along an edge to form a quadrilateral. A general configuration of $k$ non-intersecting triangles corresponds to the general N${}^k$MHV Yangian invariant shown in~\eqref{allYangian}.

\subsection{Review}

We recall that the (tree-level) amplituhedron ${\cal{A}}_{n,k}^{(m)}$ is defined~\cite{Arkani-Hamed:2013jha} as the image of the positive Grassmannian $G_+(k,n)$ under the linear map
\begin{align}\label{eq:amplituhedronmap}
C \in G_+(k,n) \quad \to \quad
Y = C \cdot Z^{\rm T} \in G(k,k+m)\,,
\end{align}
for generic positive matrix $Z \in M_+(k+m,n)$. If $\mathcal{C}$ is a positroid cell in $G_+(k,n)$, $Z(\mathcal{C})$ is its image under the amplituhedron map, and $\Omega_{\mathcal{C}}$ is the unique canonical differential form~\cite{Arkani-Hamed:2017tmz} on $G(k,k+m)$ with logarithmic singularities (only) on the boundaries of $Z(\mathcal{C})$. Then $\Omega_{\mathcal{C}}$ provides the Yangian invariant associated to $\mathcal{C}$ defined directly in the amplituhedron space, as explained in \cite{Ferro:2016zmx}. Alternatively, Yangian invariants can be represented as certain residues, or contour integrals, of the top form on $\Gr_+(k,n)$~\cite{Mason:2009qx, ArkaniHamed:2009vw}. The connection between these two ways of representing Yangian invariants is laid out in Sec.~7 of~\cite{Arkani-Hamed:2013jha}. For our purposes, the significant advantage of using the amplituhedron construction is that it enables us to write down the completely general formula for $\Omega_{\mathcal{C}}$, given below in~\eqref{allYangian}.

\begin{figure}
\begin{center}
\includegraphics[scale=0.3]{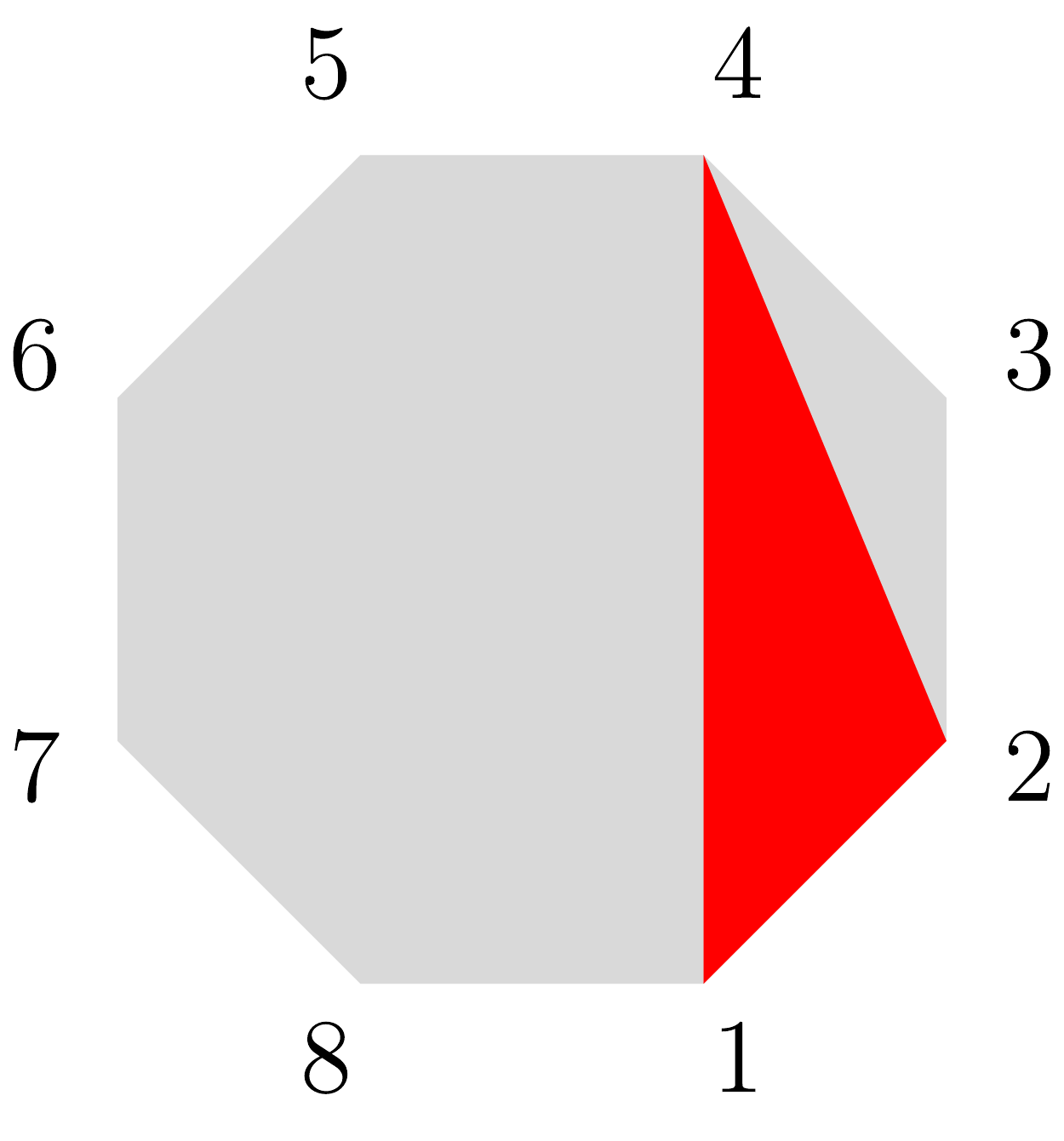}
\end{center}
\caption{Triangle $T^{(8)}_{124}$.}
\label{fig:triangle}
\end{figure}

In the remainder of this paper we specialise to $m=2$. The Yangian invariants considered in this paper correspond to $2k$-dimensional positroid cells in $G_+(k,n)$ whose images under the amplituhedron map are also $2k$-dimensional. We refer to the images of such positroid cells as \emph{generalised triangles}, and will denote them as $\mathcal{T}^{(n)}$. We will see that all Yangian invariants associated to generalised triangles can be labelled by collections of triangles in an $n$-gon. In the following we use $T^{(n)}_{abc}$ to denote the triangle with vertices $\{a,b,c\}$ inside a convex $n$-gon (see for example Fig.~\ref{fig:triangle}). We say that two triangles are {\it non-intersecting} if their interiors are disjoint, but we allow non-intersecting triangles to share an edge or a vertex.

\subsection{\texorpdfstring{$k=1$}{k=1}}

For $k=1$ we consider the most general 2-dimensional cell in $G_+(1,n)$, which can be parametrised by a positive matrix\footnote{Throughout the following we employ the unfortunately common abuse of notation by writing \emph{positive} instead of \emph{non-negative}.} of the form
\begin{equation}
C_{abc} = \bordermatrix{
&&&a&&b&&c&&\cr
&0&\ldots &\star & \ldots & \star & \ldots & \star & \ldots&0}\,,
\end{equation}
whose only non-zero entries are located in columns $\{a,b,c\}$. The image of such a cell through~\eqref{eq:amplituhedronmap} is an actual triangle with vertices $Z_a,Z_b,Z_c$ in $\mathbb{P}^2$. The corresponding Yangian invariant is~\cite{Arkani-Hamed:2013jha}
\begin{equation}\label{omegaabc}
\Omega_{abc}=\frac{\langle abc\rangle^2}{\langle Yab\rangle \langle Ybc\rangle\langle Yca\rangle}\,,
\end{equation}
and $\Omega_{abc}\, \langle Y d^2 Y\rangle$ is the canonical form with logarithmic singularities only along the three edges of the triangle, i.e.~where one of the brackets $\langle Yab\rangle$, $\langle Ybc\rangle$, or $\langle Yca\rangle$ vanishes. We introduced the following bracket notation
\begin{equation}
\langle a_1a_2\ldots a_{k+2}\rangle=\epsilon_{A_1A_2\ldots A_{k+2}}Z^{A_1}_{a_1}Z^{A_2}_{a_2}\ldots Z^{A_{k+2}}_{a_{k+2}}\,.
\end{equation}

\subsection{\texorpdfstring{$k=2$}{k=2}}

For $k=2$ we consider four-dimensional cells in $G_+(2,n)$. There are three different types of such cells, corresponding to matrix representatives that can be brought, using an appropriate $GL(2)$ transformation, to one of the following three forms:
{\let\quad\thinspace
\begin{align}
C_{a_1,b_1,c_1;a_2,b_2,c_2}& =\bordermatrix{
&&&a_1&&b_1&&c_1&&a_2&&b_2&&c_2&&\cr
&0&\ldots &\star & \ldots & \star & \ldots & \star & \ldots & 0 & \ldots & 0 & \ldots & 0& \ldots& 0\cr
&0&\ldots & 0 & \ldots & 0 &\ldots & 0 & \ldots & \star & \ldots & \star & \ldots & \star & \ldots&0 },\\\nonumber
\\
C_{a_1,b_1,c_1,d_1;a_2,b_2}& =\bordermatrix{
&&&a_1&&b_1&&c_1&&d_1&&a_2&&b_2&&\cr
&0&\ldots &\star & \ldots & \star & \ldots & \star & \ldots & \star & \ldots & 0 & \ldots & 0& \ldots& 0\cr
&0&\ldots & 0 & \ldots & 0 &\ldots & 0 & \ldots & 0 & \ldots & \star & \ldots & \star & \ldots&0 },\\\nonumber\\
C_{a_1,b_1,c_1,d_1,e_1;a_2}& =\bordermatrix{
&&&a_1&&b_1&&c_1&&d_1&&e_1&&a_2&&\cr
&0&\ldots &\star & \ldots & \star & \ldots & \star & \ldots & \star & \ldots & \star & \ldots & 0& \ldots& 0\cr
&0&\ldots & 0 & \ldots & 0 &\ldots & 0 & \ldots & 0 & \ldots & 0 & \ldots & \star & \ldots&0 }.
\end{align}}

In boundary cases some indices could be repeated (for example $c_1$ could equal $a_2$ in the first matrix). Although each type of cell is four-dimensional in $G_+(2,n)$, only the first has a four-dimensional image in $G(2,4)$; it is easy to check that cells of the second or third type have images of dimension three or two, respectively. Therefore, we are interested only in cells parametrised by matrices with three non-zero entries in columns $\{a_1,b_1,c_1\}$ in the first row and three non-zero entries in columns $\{a_2,b_2,c_2\}$ in the second row. We naturally label such a cell by a pair of triangles $T_{a_1b_1c_1}^{(n)}$ and $T_{a_2b_2c_2}^{(n)}$ inside an $n$-gon. The triangles must be non-intersecting, since otherwise the matrix $C_{a_1,b_1,c_1;a_2,b_2,c_2}$ would not be positive.

\begin{figure}[h!]
\begin{center}
\begin{tabular}{ccc}
\includegraphics[scale=0.3]{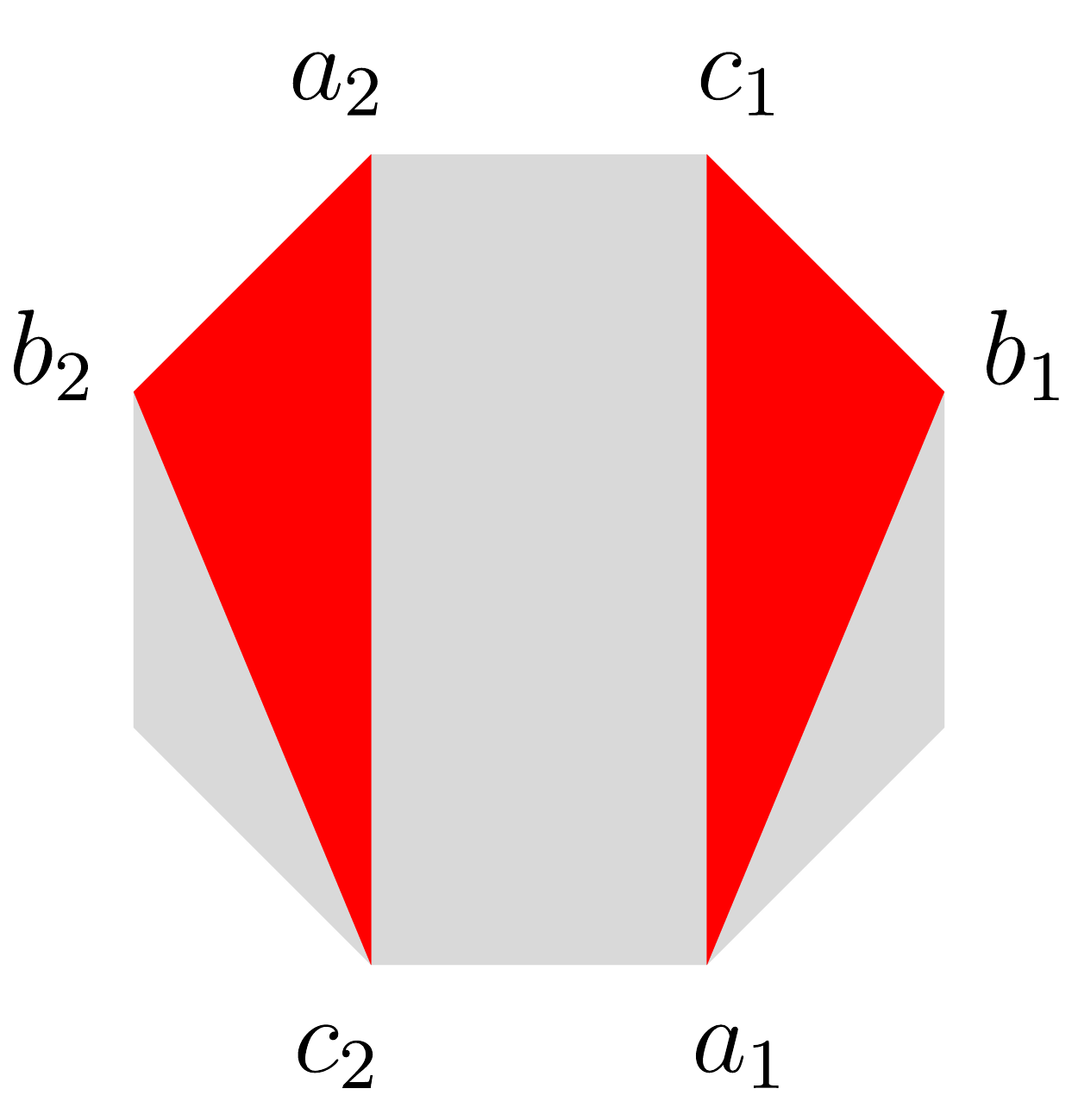}&\phantom{.}\hspace{2cm}\phantom{.}&
\includegraphics[scale=0.27]{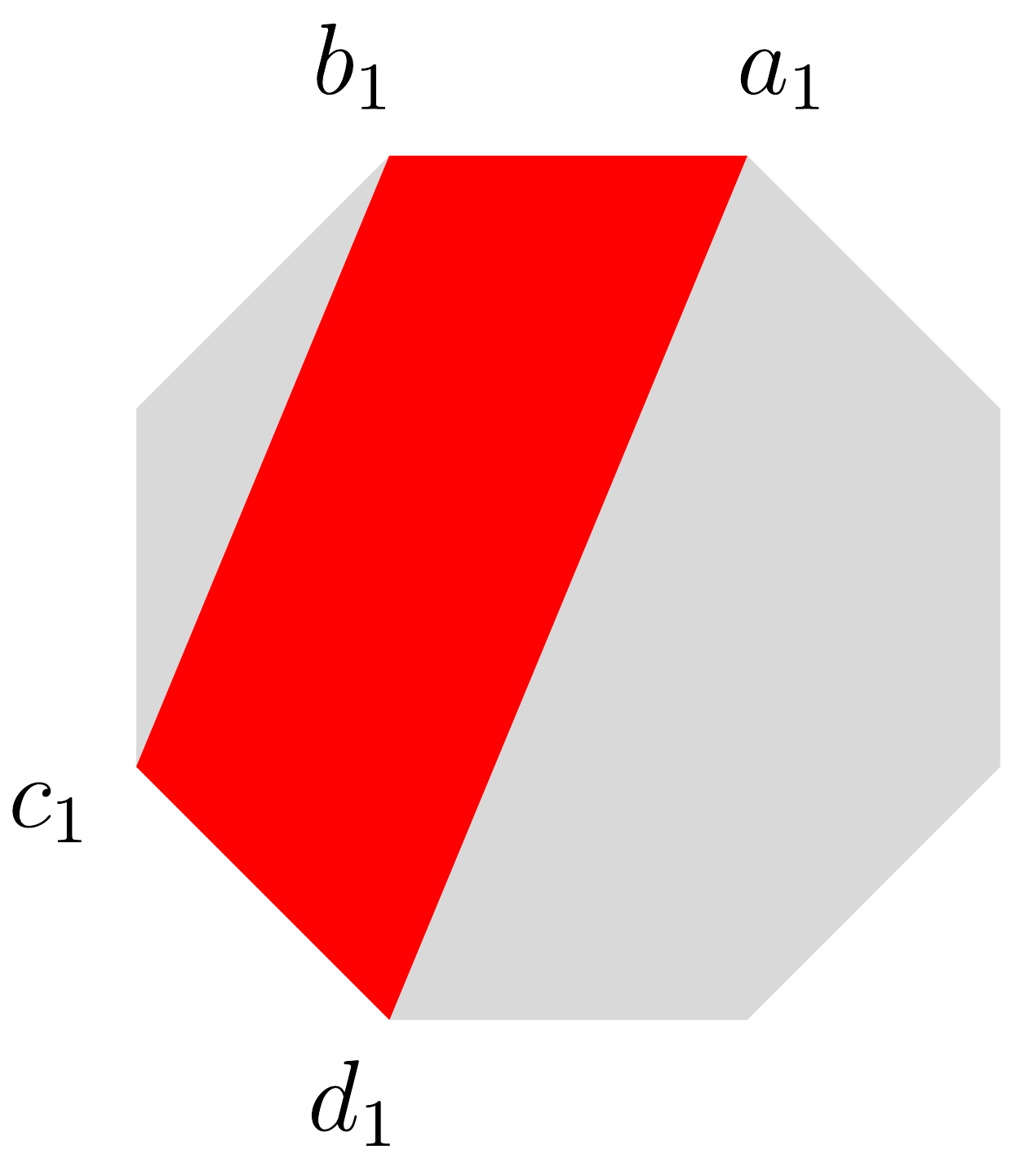}\\
a)&&b)
\end{tabular}
\end{center}
\caption{Labels for the $k=2$ generalised triangles: a) ${\mathcal{T}}_{a_1,b_1,c_1;a_2,b_2,c_2}^{(n)}$, b) ${\mathcal{T}}_{a_1,b_1,c_1,d_1}^{(n)}$.}
\label{Fig:gen.tr}
\end{figure}

There are two types of generalised triangles parametrised by $C_{a_1,b_1,c_1;a_2,b_2,c_2}$, depending on the choice of indices. If the index sets $\{a_1,b_1,c_1\}$ and $\{a_2,b_2,c_2\}$ have at most one element in common, then the two triangles $T_{a_1b_1c_1}^{(n)}$ and $T_{a_2b_2c_2}^{(n)}$ intersect at most at a single point. This type of configuration is shown in the left panel of Fig.~\ref{Fig:gen.tr}. Note that the image of this cell in $G(2,4)$, which we denote by $\mathcal{T}_{a_1,b_1,c_1; a_2b_2c_2}^{(n)}$, has six codimension-one boundaries, regardless of whether or not the triangles share a vertex.

On the other hand, the two triangles could share an edge, say $a_2 = b_1$, $b_2 = c_1$, and $d_2 = d_1$, such that they form a quadrilateral with vertices $\{a_1, b_1,c_1,d_1\}$, as shown in the right panel of Fig.~\ref{Fig:gen.tr}. The image of this cell, which we denote by $\mathcal{T}_{a_1,b_1;c_1,d_1}^{(n)}$, only has four codimension-one boundaries in $G(2,4)$. Note that there are two ways to form the same quadrilateral by joining triangles, namely $T_{a_1b_1c_1}^{(n)} \cup T_{a_1c_1d_1}^{(n)}$ or $T_{a_1b_1d_1}^{(n)} \cup T_{b_1c_1d_1}^{(n)}$. Employing a $GL(2)$ transformation, one can show that the two corresponding matrices $C_{a_1b_1c_1;a_1c_1d_1}$ and $C_{a_1b_1d_1;b_1c_1d_1}$ parametrise the same cell in $G_+(2,n)$. The fact that both the cell and its image are independent of the triangulation will become important in the following when we describe our labelling of general generalised triangles.

\begin{figure}[h]
\begin{center}
\includegraphics[scale=0.4]{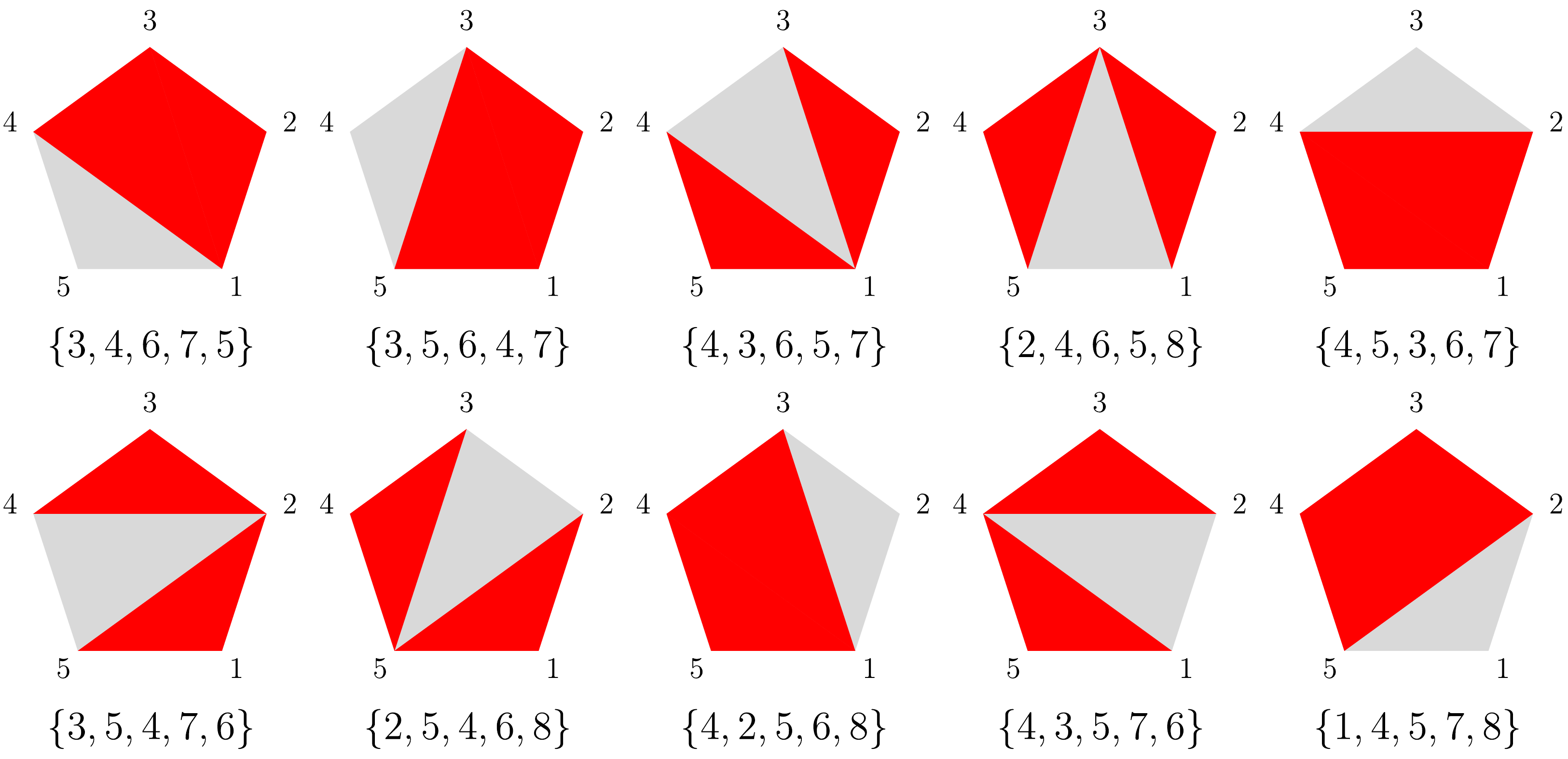}
\end{center}
\caption{The 10 generalised triangles for $n=5$, $k=2$, together with the permutations that label the corresponding positroid cells in the notation of~\cite{ArkaniHamed:2012nw}.}
\label{Fig:gen.tr.n5k2}
\end{figure}

As a concrete example, we provide the complete list of labels for generalised triangles for $k=2$, $n=5$ in Fig.~\ref{Fig:gen.tr.n5k2}.

The Yangian invariants associated to the six- and four-boundary types of generalised triangles are respectively
\begin{align}
\Omega_{a_1b_1c_1;a_2b_2c_2}&=\frac{\langle Y(a_1b_1c_1)\cap(a_2b_2c_2)\rangle^2}{\langle Ya_1b_1\rangle \langle Yb_1c_1\rangle\langle Yc_1a_1\rangle\langle Ya_2b_2\rangle \langle Yb_2c_2\rangle\langle Yc_2a_2\rangle}\,,
\label{omegaabcabc}
\\
\Omega_{a_1b_1c_1d_1}&=\frac{\langle a_1b_1c_1d_1\rangle^2}{\langle Ya_1b_1\rangle \langle Yb_1c_1\rangle\langle Yc_1d_1\rangle\langle Yd_1a_1\rangle }\,,
\label{omegaabcd}
\end{align}
where we define $\langle Y(a_1b_1c_1)\cap(a_2b_2c_2)\rangle=\langle Y_1 a_1 b_1c_1\rangle \langle Y_2a_2b_2c_2\rangle-\langle Y_2 a_1 b_1c_1\rangle \langle Y_1a_2b_2c_2\rangle$.

To summarise, the generalised triangles for $k=2$ are in one-to-one correspondence with pairs of non-intersecting triangles $T^{(n)}_{a_1 b_1 c_1}$ and $T^{(n)}_{a_2 b_2 c_2}$ that share at most a vertex, together with quadrilaterals $\lbrace a_1,b_1,c_1,d_1\rbrace$ inside an $n$-gon. Furthermore, boundaries of generalised triangles, and hence singularities of respective Yangian invariants, are in correspondence with edges of these configurations. Therefore, the set of singularities of Yangian invariants for $k=2$ correspond to a set of non-intersecting diagonals (and possible external edges) of an $n$-gon.

\subsection{General \texorpdfstring{$k$}{k}}

For general $k$ there exists a natural generalisation of the labelling we encountered for $k=1,2$. Generalised triangles are images of $2k$-dimensional positroid cells inside $G_+(k,n)$ which can be parametrised by a matrix $C_{a_1 b_1 c_1;\ldots;a_k b_k c_k}$ whose $\alpha^{\rm th}$ row has non-zero entries only in columns $\{a_\alpha,b_\alpha,c_\alpha\}$, with $\alpha=1,\ldots,k$. To this set of indices we associate the union of the $k$ non-intersecting triangles $T^{(n)}_{a_1 b_1 c_1}, \ldots, T^{(n)}_{a_k b_k c_k}$ inside an $n$-gon. If no pair of triangles share a common edge then we denote the generalised triangle associated to this configuration by $\mathcal{T}^{(n)}_{a_1b_1c_1;\ldots;a_kb_kc_k}$. If two triangles share a common edge then we combine them to form a quadrilateral. If there are more triangles sharing common edges then we remove all shared edges to further combine them into higher polygon inside the $n$-gon. Therefore, generalised triangles of the $m=2$ amplituhedron are in one-to-one correspondence with sets of non-intersecting polygons inside an $n$-gon, with no pair of polygons sharing more than a single vertex. We have checked this statement up to high values of $n$ and $k$ using the {\tt positroid} package~\cite{Bourjaily:2012gy} and we conjecture it is always true. Interestingly, we find that the intersection number (see~\cite{ArkaniHamed:2012nw, Bourjaily:2012gy}) is $\Gamma=1$ in each case. This contrasts with the situation for $m=4$ Yangian invariants, where it is not uncommon to have $\Gamma>1$.

Now it is possible to introduce a generalisation of the formulas~\eqref{omegaabc}, \eqref{omegaabcabc} and~\eqref{omegaabcd} for all Yangian invariants at general $k$. Let $s$ be the total number of polygons, and for $j=1,\ldots,s$ let $p_j$ be the number of edges of the $j^{\rm th}$ polygon. We denote the $j^{\rm th}$ polygon by $P^{(n)}_{a_{j1}, \ldots ,a_{jp_j}}$, where $a_{j1},a_{j2}, \ldots a_{jp_j}$ are its vertices. Since the total number of triangles has to be $k$ and the $j^{\rm th}$ polygon contains $p_j-2$ triangles, we must have $\sum_{j=1}^s (p_j-2)=k$. Then the Yangian invariant associated to the collection of polygons $\lbrace P^{(n)}_{a_{11},\ldots,a_{1p_1}},\ldots,P^{(n)}_{a_{s1},\ldots,a_{sp_s}} \rbrace$ is
\begin{equation}\label{allYangian}
\Omega_{a_{11},\ldots,a_{1p_1};\ldots;a_{s1},\ldots,a_{sp_s}}=\frac{\langle Y (a_{11}\ldots a_{1p_1}) \cap\ldots\cap (a_{s1}\ldots a_{sp_s}) \rangle^2}{(\langle Y a_{11} a_{12}\rangle\ldots\langle Y a_{1p_1} a_{11}\rangle)\ldots(\langle Y a_{s1} a_{s2}\rangle\ldots\langle Y a_{sp_s} a_{s1}\rangle)}\,,
\end{equation}
where we have defined
\begin{multline}
\langle Y (a_{11}\ldots a_{1p_1}) \cap\ldots\cap (a_{s1}\ldots a_{sp_s}) \rangle
\\
= \epsilon_{\alpha^1_1\ldots\alpha^s_{p_s-2}} \langle Y^\perp_{\alpha^1_1\ldots\alpha^1_{p_1-2}} a_{11}\ldots a_{1p_1} \rangle\ldots\langle Y^\perp_{\alpha^s_{1}\ldots\alpha^s_{p_s-2}} a_{s1}\ldots a_{sp_s} \rangle\,,
\end{multline}
and $Y^\perp_{\alpha_1\ldots \alpha_r}$ parametrises the orthogonal complement of a collection of $Y$'s,
\begin{equation}
Y^\perp_{\alpha_1\ldots\alpha_r}=\epsilon_{\alpha_1\ldots\alpha_k} Y_{\alpha_{r+1}} \wedge \ldots \wedge Y_{\alpha_{k}}\,.
\end{equation}

We note that if $p_j=3$ for all $j=1,\ldots,s$, i.e.~all polygons are triangles, then~\eqref{allYangian} agrees with~\cite{Arkani-Hamed:2017tmz}\footnote{Notice a typo in the numerator of formula (7.52) in~\cite{Arkani-Hamed:2017tmz}: the power should be $2$ instead of $k$.}. Moreover, the formula~\eqref{allYangian} nicely encodes the geometry of the set of polygons of our labels. The building blocks of the numerator are $s$ brackets, one for each polygon, with the $j^{\rm th}$ bracket containing all the vertices of the $j^{\rm th}$ polygon. Meanwhile the denominator has exactly $s$ factors, and the $j^{\rm th}$ factor is a product of brackets of the type $\langle Y ab \rangle$ over all edges $\lbrace a,b \rbrace$ of the $j^{\rm th}$ polygon. When the $i^{\rm th}$ and $j^{\rm th}$ polygon share an edge $\lbrace a, b\rbrace$, so that they combine to form a $(p_i+p_j-2)$-gon, then formula~\eqref{allYangian} nicely rearranges to give the Yangian invariant associated with the same set of polygons, but with the $i^{\rm th}$ and $j^{\rm th}$ polygon replaced by the merged $(p_i+p_j-2)$-gon. In particular, it can be shown that the numerator factorises and one gets an overall factor of $\langle Yab\rangle^2$, which cancels the singularities associated with the shared edge $\lbrace a, b\rbrace$ from the denominator, as expected.

To summarise, we see that in general, all singularities of a given Yangian invariant correspond to a subset of non-intersecting diagonals inside of an $n$-gon (and possibly external edges). In the appendix we provide an explicit enumeration of the number of $n$-particle N${}^k$MHV Yangian invariants for $m=2$.

\section{\texorpdfstring{$A_{n-3}$}{A(n-3)} Cluster Adjacency}
\label{sec:3}

Tree-level \emph{cluster adjacency} in $\mathcal{N}=4$ Yang-Mills theory is the conjectured~\cite{Drummond:2018dfd} property that every (rational) Yangian invariant has poles given by some collection of $\mathcal{A}$-coordinates of the $\Gr(4,n)$ cluster algebra that can be found together in common cluster. So far, evidence supporting this conjecture is restricted to relatively small $n$ and $k$. In~\cite{Drummond:2018dfd} several examples were checked by explicitly identifying a suitable cluster for several relatively simple Yangian invariants. Later in~\cite{Mago:2019waa} a computationally efficient method (first explained in~\cite{Golden:2019kks}) for testing whether two cluster coordinates belong in a common cluster was used to provide further evidence for this conjecture for various somewhat higher $n$ and $k$.

The natural generalisation of the cluster adjacency conjecture for general $m$ would posit that the poles of every Yangian invariant are given by some subset of the $\mathcal{A}$-coordinates of some cluster of the $\Gr(m,n)$ algebra. For $m=2$ this is the same as the classic $A_{n-3}$ algebra, whose structure is completely understood~\cite{FZ1}. This algebra has $\frac{1}{n-1} \binom{2n-4}{n-2}$ clusters, each containing $2n{-}3$ $\mathcal{A}$-coordinates. These numbers are respectively the number of distinct triangulations of an $n$-gon, and the number of edges (including external edges) in any such triangulation. If we label the vertices of a regular $n$-gon by homogeneous coordinates $z_1, \ldots, z_n$ of $n$ points in $\mathbb{P}^1$ and let $\langle a\,b \rangle = \epsilon_{AB}\,z_a^A z_b^B$, as usual, then the coordinates are enumerated as follows. Each cluster contains the $n$ coordinates $\langle 1\,2\rangle, \langle 2\,3\rangle, \ldots, \langle n\,1 \rangle$ corresponding to the external edges, together with precisely $n{-}3$ additional $\langle a\,b\rangle$'s that correspond to the internal edges of the triangulation.

In light of this discussion it is now essentially obvious that every Yangian invariant in the $m=2$ version of $\mathcal{N}=4$ Yang-Mills theory, whose generic form is shown in~\eqref{allYangian}, manifestly satisfies cluster adjacency. The only detail requiring comment is that, as discussed in~\cite{Arkani-Hamed:2013jha}, it is possible to replace the $(k+2)$-component $Z_a$'s appearing in a bracket of the form $\langle Y a b \rangle$ by their projections to the ``ordinary'' two-component homogeneous coordinate $z_a$ on $\mathbb{P}^1$, by setting
\begin{align}
Y = \left( \begin{matrix}
0_{2 \times k} \\
1_{k \times k} \end{matrix}\right).
\end{align}
That is, as far as the denominator of~\eqref{allYangian} is concerned, we can simply replace every $\langle Y a b \rangle$ by $\langle a\, b\rangle$, making the cluster adjacency manifest. The numerator of a Yangian invariant will in general be a rather non-trivial polynomial in the $z_a$'s and their Grassmann partners, but the numerator is of no concern to us.

The fact that the connection between cluster adjacency and Yangian invariants is strikingly simple for $m=2$ provides some circumstantial evidence in support of our hope that the same will be true for the apparently much more non-trivial case of $m=4$ (or perhaps even for general $m$). It also gives support to the suggestion made in~\cite{Mago:2019waa} that the connection between cluster adjacency and Yangian invariants might admit a mathematical explanation that is independent of the physics of scattering amplitudes, and most likely originates from the geometry of the amplituhedron.

If there is to be, one day, an analytic proof of the tree-level cluster adjacency conjecture, it is natural to speculate that it may hinge on the fact that it is known~\cite{Drummond:2010uq} that every positive Yangian invariant can be written as Grassmannian integral~\cite{ArkaniHamed:2009dn} (more specifically, over the momentum twistor Grassmannian~\cite{Mason:2009qx,ArkaniHamed:2009vw}) over a contour associated to a positroid cell. Here, then, we have access to relatively simple situations in which both the \emph{integrand} (the natural top form on $\Gr(k,n)$) and the \emph{integral} (the resulting Yangian invariant) are both \emph{rational functions}.  It would be extremely exciting to learn what property of the former is responsible, after integration, for cluster adjacency of the latter.  This could help point the way towards answering the long-standing, but much more complicated, question of how the cluster structure of integrands~\cite{ArkaniHamed:2012nw} in SYM theory is related, upon integration, to the cluster structure of the resulting polylogarithmic functions that appear in amplitudes.

\begin{table}[!ht]
\begin{center}
\begin{tabular}{c|cccccccc}
\backslashbox{$k$}{$n$}& 3&4&5&6&7&8&9&10\\
\hline
0&1&1&1&1&1&1&1&1\\
\hline
1&1&4&10&20&35&56&84&120\\
\hline
2&0&1&10&48&161&434&1008&2100\\
\hline
3&0&0&1&20&161&824&3186&10152\\
\hline
4&0&0&0&1&35&434&3186&16840\\
\hline
5&0&0&0&0&1&56&1008&10152\\
\hline
6&0&0&0&0&0&1&84&2100\\
\hline
7&0&0&0&0&0&0&1&120\\
\hline
8&0&0&0&0&0&0&0&1\\
\hline
\bf Total&\bf 2&\bf 6&\bf 22&\bf 90&\bf 394&\bf 1806&\bf 8558&\bf 41586\\
\end{tabular}
\end{center}
\caption{The number of generalised triangles for $n<11$.}
\label{table}
\end{table}

\acknowledgments

This work was supported in part by the US Department of Energy under contract {DE}-{SC}0010010 Task A (MS, AV) and by Simons Investigator Award \#376208 (AV). In addition MP would like to thank `Fondazione A. Della Riccia' for financial support, and MS and AV thank the CERN Theory Group for hospitality during the completion of this work. This work was performed in part at Aspen Center for Physics, which is supported by National Science Foundation grant PHY-1607611. TL was partially supported by a grant from the Simons Foundation. 

\appendix

\section{Number of Generalised Triangles}

We have tabulated the number of generalised triangles for $n<11$ In Tab.~\ref{table}. Reading down the columns gives $1,1,1,1,4,1,1,10,10,1,1,20,48,20,1,1,35,161,161,35,\ldots$ which is sequence A175124 in~\cite{OEIS}. It is generated by the coefficients of the inverse series of $\frac{x(1-p q x^2)}{(1+px)(1+qx)}$. The sequence of the total number of generalised triangles for given $n$: $2,6,22,90,394,\ldots$, is known as the sequence of large Schr\"oder numbers. Interestingly, the definition of Schr\"oder numbers as the number of all configurations of non-intersecting triangles in an $n$-gon, seems to be absent in the literature.

\bibliographystyle{JHEP}

\bibliography{cluster_yangians}

\providecommand{\href}[2]{#2}\begingroup\raggedright\begin{thebibliography}{10}

\bibitem{FZ1}
S.~Fomin and A.~Zelevinsky, \emph{{Cluster Algebras I: Foundations}},
  {\emph{Journal of the American Mathematical Society} {\bfseries 15} (2002)
  497}.

\bibitem{GSV}
M.~Gekhtman, M.~Z. Shapiro and A.~D. Vainshtein, \emph{Cluster algebras and
  poisson geometry}, {\emph{Moscow Mathematical Journal} {\bfseries 3} (2003)
  899} [\href{https://arxiv.org/abs/math/0208033}{{\ttfamily math/0208033}}].

\bibitem{scott2006grassmannians}
J.~S. Scott, \emph{Grassmannians and cluster algebras}, {\emph{Proceedings of
  the London Mathematical Society} {\bfseries 92} (2006) 345}
  [\href{https://arxiv.org/abs/math/0311148}{{\ttfamily math/0311148}}].

\bibitem{Golden:2013xva}
J.~Golden, A.~B. Goncharov, M.~Spradlin, C.~Vergu and A.~Volovich,
  \emph{{Motivic Amplitudes and Cluster Coordinates}},
  \href{https://doi.org/10.1007/JHEP01(2014)091}{\emph{JHEP} {\bfseries 01}
  (2014) 091} [\href{https://arxiv.org/abs/1305.1617}{{\ttfamily 1305.1617}}].

\bibitem{Golden:2014xqa}
J.~Golden, M.~F. Paulos, M.~Spradlin and A.~Volovich, \emph{{Cluster
  Polylogarithms for Scattering Amplitudes}},
  \href{https://doi.org/10.1088/1751-8113/47/47/474005}{\emph{J. Phys.}
  {\bfseries A47} (2014) 474005}
  [\href{https://arxiv.org/abs/1401.6446}{{\ttfamily 1401.6446}}].

\bibitem{Hodges:2009hk}
A.~Hodges, \emph{{Eliminating spurious poles from gauge-theoretic amplitudes}},
  \href{https://doi.org/10.1007/JHEP05(2013)135}{\emph{JHEP} {\bfseries 05}
  (2013) 135} [\href{https://arxiv.org/abs/0905.1473}{{\ttfamily 0905.1473}}].

\bibitem{DelDuca:2016lad}
V.~Del~Duca, S.~Druc, J.~Drummond, C.~Duhr, F.~Dulat, R.~Marzucca et~al.,
  \emph{{Multi-Regge kinematics and the moduli space of Riemann spheres with
  marked points}}, \href{https://doi.org/10.1007/JHEP08(2016)152}{\emph{JHEP}
  {\bfseries 08} (2016) 152}
  [\href{https://arxiv.org/abs/1606.08807}{{\ttfamily 1606.08807}}].

\bibitem{Arkani-Hamed:2017tmz}
N.~Arkani-Hamed, Y.~Bai and T.~Lam, \emph{{Positive Geometries and Canonical
  Forms}}, \href{https://doi.org/10.1007/JHEP11(2017)039}{\emph{JHEP}
  {\bfseries 11} (2017) 039}
  [\href{https://arxiv.org/abs/1703.04541}{{\ttfamily 1703.04541}}].

\bibitem{karp2017decompositions}
S.~N. Karp, L.~K. Williams and Y.~X. Zhang, \emph{Decompositions of
  amplituhedra},  \href{https://arxiv.org/abs/1708.09525}{{\ttfamily
  1708.09525}}.

\bibitem{Galashin:2018fri}
P.~Galashin and T.~Lam, \emph{{Parity duality for the amplituhedron}},
  \href{https://arxiv.org/abs/1805.00600}{{\ttfamily 1805.00600}}.

\bibitem{Ferro:2018vpf}
L.~Ferro, T.~{\L}ukowski and M.~Parisi, \emph{{Amplituhedron meets
  Jeffrey–Kirwan residue}},
  \href{https://doi.org/10.1088/1751-8121/aaf3c3}{\emph{J. Phys.} {\bfseries
  A52} (2019) 045201} [\href{https://arxiv.org/abs/1805.01301}{{\ttfamily
  1805.01301}}].

\bibitem{Lukowski:2019kqi}
T.~{\L}ukowski, \emph{{On the Boundaries of the m=2 Amplituhedron}},
  \href{https://arxiv.org/abs/1908.00386}{{\ttfamily 1908.00386}}.

\bibitem{Drummond:2018dfd}
J.~Drummond, J.~Foster and {\" O}.~G{\"u}rdo{\u g}an, \emph{{Cluster adjacency
  beyond MHV}}, \href{https://doi.org/10.1007/JHEP03(2019)086}{\emph{JHEP}
  {\bfseries 03} (2019) 086}
  [\href{https://arxiv.org/abs/1810.08149}{{\ttfamily 1810.08149}}].

\bibitem{Mago:2019waa}
J.~Mago, A.~Schreiber, M.~Spradlin and A.~Volovich, \emph{{Yangian Invariants
  and Cluster Adjacency in $\mathcal{N}=4$ Yang-Mills}},
  \href{https://arxiv.org/abs/1906.10682}{{\ttfamily 1906.10682}}.

\bibitem{Drummond:2017ssj}
J.~Drummond, J.~Foster and {\" O}.~G{\"u}rdo{\u g}an, \emph{{Cluster Adjacency
  Properties of Scattering Amplitudes in $\mathcal{N}=4$ Supersymmetric
  Yang-Mills Theory}},
  \href{https://doi.org/10.1103/PhysRevLett.120.161601}{\emph{Phys. Rev. Lett.}
  {\bfseries 120} (2018) 161601}
  [\href{https://arxiv.org/abs/1710.10953}{{\ttfamily 1710.10953}}].

\bibitem{ArkaniHamed:2012nw}
N.~Arkani-Hamed, J.~L. Bourjaily, F.~Cachazo, A.~B. Goncharov, A.~Postnikov and
  J.~Trnka, \emph{{Grassmannian Geometry of Scattering Amplitudes}}. Cambridge
  University Press, 2016,
  \href{https://doi.org/10.1017/CBO9781316091548}{10.1017/CBO9781316091548},
  [\href{https://arxiv.org/abs/1212.5605}{{\ttfamily 1212.5605}}].

\bibitem{Arkani-Hamed:2013jha}
N.~Arkani-Hamed and J.~Trnka, \emph{{The Amplituhedron}},
  \href{https://doi.org/10.1007/JHEP10(2014)030}{\emph{JHEP} {\bfseries 10}
  (2014) 030} [\href{https://arxiv.org/abs/1312.2007}{{\ttfamily 1312.2007}}].

\bibitem{Mason:2009qx}
L.~J. Mason and D.~Skinner, \emph{{Dual Superconformal Invariance, Momentum
  Twistors and Grassmannians}},
  \href{https://doi.org/10.1088/1126-6708/2009/11/045}{\emph{JHEP} {\bfseries
  11} (2009) 045} [\href{https://arxiv.org/abs/0909.0250}{{\ttfamily
  0909.0250}}].

\bibitem{ArkaniHamed:2009vw}
N.~Arkani-Hamed, F.~Cachazo and C.~Cheung, \emph{{The Grassmannian Origin Of
  Dual Superconformal Invariance}},
  \href{https://doi.org/10.1007/JHEP03(2010)036}{\emph{JHEP} {\bfseries 03}
  (2010) 036} [\href{https://arxiv.org/abs/0909.0483}{{\ttfamily 0909.0483}}].

\bibitem{ArkaniHamed:2009dg}
N.~Arkani-Hamed, J.~Bourjaily, F.~Cachazo and J.~Trnka, \emph{{Unification of
  Residues and Grassmannian Dualities}},
  \href{https://doi.org/10.1007/JHEP01(2011)049}{\emph{JHEP} {\bfseries 01}
  (2011) 049} [\href{https://arxiv.org/abs/0912.4912}{{\ttfamily 0912.4912}}].

\bibitem{ArkaniHamed:2009sx}
N.~Arkani-Hamed, J.~Bourjaily, F.~Cachazo and J.~Trnka, \emph{{Local Spacetime
  Physics from the Grassmannian}},
  \href{https://doi.org/10.1007/JHEP01(2011)108}{\emph{JHEP} {\bfseries 01}
  (2011) 108} [\href{https://arxiv.org/abs/0912.3249}{{\ttfamily 0912.3249}}].

\bibitem{Drummond:2010uq}
J.~M. Drummond and L.~Ferro, \emph{{The Yangian origin of the Grassmannian
  integral}}, \href{https://doi.org/10.1007/JHEP12(2010)010}{\emph{JHEP}
  {\bfseries 12} (2010) 010} [\href{https://arxiv.org/abs/1002.4622}{{\ttfamily
  1002.4622}}].

\bibitem{Ashok:2010ie}
S.~K. Ashok and E.~Dell'Aquila, \emph{{On the Classification of Residues of the
  Grassmannian}}, \href{https://doi.org/10.1007/JHEP10(2011)097}{\emph{JHEP}
  {\bfseries 10} (2011) 097} [\href{https://arxiv.org/abs/1012.5094}{{\ttfamily
  1012.5094}}].

\bibitem{Drummond:2010qh}
J.~M. Drummond and L.~Ferro, \emph{{Yangians, Grassmannians and T-duality}},
  \href{https://doi.org/10.1007/JHEP07(2010)027}{\emph{JHEP} {\bfseries 07}
  (2010) 027} [\href{https://arxiv.org/abs/1001.3348}{{\ttfamily 1001.3348}}].

\bibitem{Ferro:2016zmx}
L.~Ferro, T.~{\L}ukowski, A.~Orta and M.~Parisi, \emph{{Yangian symmetry for
  the tree amplituhedron}},
  \href{https://doi.org/10.1088/1751-8121/aa7594}{\emph{J. Phys.} {\bfseries
  A50} (2017) 294005} [\href{https://arxiv.org/abs/1612.04378}{{\ttfamily
  1612.04378}}].

\bibitem{ArkaniHamed:2010gg}
N.~Arkani-Hamed, J.~L. Bourjaily, F.~Cachazo, A.~Hodges and J.~Trnka, \emph{{A
  Note on Polytopes for Scattering Amplitudes}},
  \href{https://doi.org/10.1007/JHEP04(2012)081}{\emph{JHEP} {\bfseries 04}
  (2012) 081} [\href{https://arxiv.org/abs/1012.6030}{{\ttfamily 1012.6030}}].

\bibitem{Bourjaily:2012gy}
J.~L. Bourjaily, \emph{{Positroids, Plabic Graphs, and Scattering Amplitudes in
  Mathematica}},  \href{https://arxiv.org/abs/1212.6974}{{\ttfamily
  1212.6974}}.

\bibitem{Golden:2019kks}
J.~Golden, A.~J. McLeod, M.~Spradlin and A.~Volovich, \emph{{The Sklyanin
  Bracket and Cluster Adjacency at All Multiplicity}},
  \href{https://doi.org/10.1007/JHEP03(2019)195}{\emph{JHEP} {\bfseries 03}
  (2019) 195} [\href{https://arxiv.org/abs/1902.11286}{{\ttfamily
  1902.11286}}].

\bibitem{ArkaniHamed:2009dn}
N.~Arkani-Hamed, F.~Cachazo, C.~Cheung and J.~Kaplan, \emph{{A Duality For The
  S Matrix}}, \href{https://doi.org/10.1007/JHEP03(2010)020}{\emph{JHEP}
  {\bfseries 03} (2010) 020} [\href{https://arxiv.org/abs/0907.5418}{{\ttfamily
  0907.5418}}].

\bibitem{OEIS}
N.~J. Sloane et~al., \emph{The on-line encyclopedia of integer sequences},
  2003--.

\end{thebibliography}\endgroup

\end{document}